\documentclass{amsart}
\begin{document}
\title{Warped product approach to universe with non-smooth scale factor}
\author{Jaedong Choi, Soon-Tae Hong}
\address{\noindent Jaedong choi Department of Mathematics P.O. Box 335-2 Air Force Academy,
Ssangsu, Namil, Cheongwon, Chungbuk, 363-849 Korea}
\email{\noindent jdong@afa.ac.kr}
\address{\noindent Soon-Tae Hong Department of Science Education,
Ewha Womans University, Seoul 120-750 Korea} \email{\noindent
soonhong@ewha.ac.kr} \noindent
\begin{abstract}
In the framework of Lorentzian warped products, we study the Friedmann-Robertson-Walker
cosmological model to investigate non-smooth curvatures
associated with multiple discontinuities involved in the
evolution of the universe.  In particular we analyze non-smooth features of the spatially 
flat Friedmann-Robertson-Walker universe by introducing double discontinuities occurred at the
radiation-matter and matter-lambda phase transitions in astrophysical phenomenology.
\end{abstract}
\vskip10pt
\maketitle

\noindent{\bf I. Introduction} \vskip10pt

Since the cosmic microwave background was discovered, there have been many ideas and proposals to figure out how the universe has evolved.  The standard big bang cosmological model based on the Friedmann-Robertson-Walker (FRW) spacetimes has led to the inflationary cosmology~\cite{guth} and nowadays to the M-theory cosmology with bouncing universes~\cite{seiberg}.  These spacetimes are foliated by a special set of spacelike hypersurfaces such that each hypersurface corresponds to an instant of time. From a physical point of view, these warped product spacetimes are interesting since they include classical examples of spacetime such as the FRW manifold and the intermediate zone of Reissner-Nordstr\"{o}m (RN) manifold~\cite{rn,ksy}.\vskip5pt 

The Lorentzian manifolds with non-smooth metric tensors have been extensively
discussed from various view points~\cite{choi00,fs,li,sm1,sm2}.  In a spacetime where the metric tensor 
is continuous but has a jump in its first and second derivatives across a submanifold in an admissible 
coordinate system, one can have a curvature tensor containing a 
Dirac delta function~\cite{gk12}.  The support of this
distribution may be of three, two, or one dimensional or may even consist of a single event.  Moreover, 
Lichnerowicz's formalism~\cite{li} for dealing with such tensors is modified so that one can 
obtain the Riemannian curvature tensor 
and Ricci curvature tensor defined in the sense of distributions.\vskip5pt  

A general theory for matching two solutions of the Einstein field equations has been proposed~\cite{sm1,sm2} at arbitrary shock-wave interface across which the metric tensor is $C^0$-Lorentzian, namely  at smooth surface across which the first
derivatives of the metric suffer at worst a jump discontinuity, so that  the simplest solution of Einstein equations can incorporate a shock-wave into a standard FRW metric whose equation of state accounts for the Hubble constant and the microwave background radiation temperature.  There have been later presented the evolution of the one point probability distribution function of the cosmological density field based on an exact statistical treatment~\cite{ta}.\vskip5pt 

On the other hand, the concept of a warped product manifold was introduced by Bishop and O'Neill long ago~\cite{bo}, and
it was later connected to general relativity~\cite{bep} and semi-Riemannian geometry~\cite{on} by elevating warped products to a central role.  Warped product spaces has been also extended to a richer class of spaces involving multiply product spaces~\cite{choi00,fs}.  One of us has investigated the curvature of a multiply warped product possessing $C^0$-warping functions with a discontinuity at a single point~\cite{choi00}, and in this paper we will generalize this 
result to a warped product spacetime with multiple discontinuities associated with cosmological phenomenology.  
Of particular interest are spacetimes with metric tensors which fail to be $C^{1}$ across multiple points on the hypersurface, and is $C^{\infty}$ off the hypersurface. We will also study the Lorentzian metric which fails to be $C^{0}$ across multiple points on the 
hypersurfaces and is $C^{\infty}$ off the hypersurfaces.\vskip5pt  

In this paper, as a cosmological model we will exploit the FRW spacetimes ${M_0}\times_f H$, which can 
be treated as a warped product manifold possessing warping function (or scale factor) $f$ with time 
dependence, to investigate the non-smooth curvature associated
with the multiple discontinuities involved in the evolution of the
universe. We will also analyze non-smooth features
of the spatially flat FRW universe by introducing double
discontinuities occurred at the radiation-matter and matter-lambda
phase transitions in the astrophysical phenomenology.\vskip5pt  

In section 2 we will introduce the warped product spacetime with multiple warping functions 
and extend the warped product scheme to the case with  multiple discontinuities in the FRW 
cosmological model in section 3.  We will study the realistic cosmological 
phenomenology in the spatially flat FRW universe associated with the radiation-matter and 
matter-lambda phase transitions in section 4.\vskip10pt

\noindent{\bf II.  Warped product spacetime with multiple warping functions}\vskip10pt

In this section we briefly recapitulate the curvature of the warped product approach to spacetime with multiple $C^0$-warping functions at a single point.
\vskip10pt
 
{\bf Definition 2.1}~\cite{choi00}\ \ A multiply warped products spacetime with base $(M_0,-dt^2)$, fibers $(F_i,g_i)$ $i=1,...,n$ and warping functions $f_i>0$ is the product manifold $({M_0}\times F_1\times\cdots\times F_n, g)$ endowed 
with the Lorentzian metric:
$$
g=-\pi_{M_0}^{\ast}dt^2+\sum_{i=1}^n(f_i\circ\pi_{M_0})^2\pi_i^\ast g_i
$$ 
where $\pi_{M_0}$, $\pi_i$\ $(i=1,...,n)$ are the natural projections of 
${M_0}\times F_1\times\cdots\times F_n$ onto ${M_0}$ and $F_1$,...,$F_n$, respectively.  
For a specific case of $M_{0}=R$ and $g_{M_{0}}=-dt^{2}$, the Lorentzian metric is given by
$$
g=-dt^{2}+\sum_{i=1}^{n}f_{i}^{2}g_{i}.
$$\vskip10pt

{\bf Proposition 2.2}~\cite{choi00}\ \ Let $M=M_0\times_{f_1} F_1\times\cdots\times_{f_n} F_n$ be a multiply 
warped products with Riemannian curvature tensor $R$. If $X$, $Y\in \frak{L}(M_0)$, $U_i$, 
$V_i$, $W_i\in\frak{L}(F_i)$ ($n=1,2,...,n$), $f_i\in C^0(S)$ at a single point 
$p\in M_{0}$, and $S=\{p\}\times _{f_1} F_1\times\cdots\times_{f_n} F_n$, then
\begin{eqnarray}
&(i)&R_{XU_i}{U_j}=R_{U_iU_j}{X}=R_{{U_j}{X}}{U_i}=0\ {\text{for}}\ \  
 i\not=j,\nonumber\\
&(ii)&R_{U_iX}Y = U_{i}X^{1}Y^{1}\frac{f_i''(t)+\delta(t-p)({f_i'}^{+}-{f_i'}^{-})}{f_i},\nonumber\\
&(iii)&R_{XU_i}{U_j}=R_{U_iU_j}{X}=R_{{U_j}{X}}{U_i}=0,~~{\text{for}}\ \  
 i\not=j,\nonumber\\
&(iv)&R_{XY}U_i=0,~~{\rm for}\ \ i=1,..,n ,\nonumber\\
&(v)&R_{U_iV_i}U_j=0,~~{\rm for}\ \ i\not=j,\nonumber\\
&(vi)&R_{U_iU_j}V_j=U_i\langle U_j,V_j\rangle
\frac{({{f_i}'}^{+}+{{f_i}'}^{-})({{f_j}'}^{+}+{{f_j}'}^{-})}{f_{i}f_{j}},
~~{\text{for}}\ \  i\not=j,\nonumber\\
&(vii)&R_{U_iV_i}W_i={}^{F_i}R_{U_iV_i}W_i \nonumber\\
&&~~~~~~~+(\langle U_i, W_i\rangle V_i -\langle V_i, W_i\rangle U_i)
\frac{{f_i'}^{+}\mu(t-p)+{f_i'}^{-}\mu(p-t)}{{f_i}^2},\nonumber
\end{eqnarray}
where $X=X^{1}\partial/\partial_{t}$ and
$Y=Y^{1}\partial/\partial_{t}$, and $\mu(t-p)$ and $\delta(t-p)$ are the unit step
function and the delta function, respectively.\vskip10pt

{\bf Proposition 2.3}~\cite{choi00}\ \ Let $M=M_0\times_{f_1} F_1\times...\times_{f_n} F_n$ 
be a multiply warped products with Riemannian curvature tensor $R$. If $X$, 
$Y\in \frak{L}(M_0)$, $U_i$, $V_i\in\frak{L}(F_i)$ ($n=1,2,...,n$), $d_i={\rm dim}~F_i$, 
$f_i\in C^0(S)$ at a single point 
$p\in M_{0}$, and $S=\{p\}\times _{f_1} F_1\times\cdots\times_{f_n} F_n$, then  
\begin{eqnarray} 
&(i)&~{\rm Ric}(X,Y)=-\sum_{i=1}^{n}d_i
X^{1}Y^{1}\frac{f_i''(t)+\delta(t-p)
\left({f_i'}^{+}-{f_i'}^{-}\right)}{f_i},\nonumber\\
&(ii)&~{\rm Ric}(X,U_i)=0,\nonumber\\
&(iii)&~{\rm Ric}(U_i,V_i)={}^{F_{i}}{\rm Ric}(U_i, V_i)+\langle
U_i, V_i\rangle
\frac{f_i''(t)+\delta(t-p)({f_i'}^{+}-{f_i'}^{-})}{f_i}
\nonumber\\
&&~~~+\langle U_i, V_i\rangle\left[
(d_{i}-1)\frac{{f_{i}'}^{+}-{f_{i}'}^{-}}{f_{i}^{2}}
+\sum_{j\neq i} d_{j}\frac{\langle {f_i'}^{+}-{f_i'}^{-},\ {f_j'}^{+}
-{f_j'}^{-}\rangle}{f_i f_j}\right],\nonumber\\
&(iv)&~{\rm Ric}(U_i, U_j)=0,~~{\rm for}~~i\neq j,\nonumber
\end{eqnarray}
where $X=X^{1}\partial/\partial_{t}$ and
$Y=Y^{1}\partial/\partial_{t}$, and $\delta(t-p)$ is the delta function.\vskip10pt

\noindent{\bf III. FRW metric with multiple discontinuities}\vskip10pt

The FRW spacetime is one of  the {\it warped product} manifold where the base is an open interval $M_0$ of $R$ with usual metric reversed $(M_0, -dt^2)$, the fiber is a 3-dimensional Riemannian manifold $(F, g_F)$ and the warping function $f$ is any positive function $f$ on $M_0$. The Robertson-Walker spacetime is then the product manifold $M=M_0\times_f H$ 
endowed with the Lorentzian
metric $g=-dt^2+f^2(t)g_H$ with $f$ being the scale factor of the FRW universe associated with universal expansion. This warping function $f$ is a function of time alone and it measures how physical separations change with time. The dynamics of the expanding universe only appears implicitly in the time dependence of the warping function (or scalar factor) $f$.\vskip5pt

Consider the spacetime $(M,\ g)$ with metric
$g=-dt^2+f^2d\sigma^2$ in the form of warped products. Let
$M=M_0\times_{f} H$ be a warped product with $g_{M_0}=-dt^2$. Let $f>0$ be smooth functions on $M_0=(t_{0},\ t_{\infty})$. Assume
$f\in C^{\infty}$ for $t\not=t_{i}$ and $f\in C^1$ at $t=t_{i}$
$(i=1,2,...,n)$.  When $f\in C^1$ at points $t_{i}\in (t_{0},\
t_{\infty})$ and $S=\{t_{i}\}\times_{f}H$, we define $f\in
C^{1}(S)$ as a collection of functions $\{f^{(i)}\}$ with
$f^{(i)}$ piecewisely defined on the intervals $t_{i}\leq t\leq
t_{i+1}$ $(i=0,1,2,...,n)$ with $t_{n+1}=t_{\infty}$. Since $f\in
C^{1}(S)$, we have $f^{(i-1)}=f^{(i)}$, ${f^{(i-1)}}'={f^{(i)}}'$
but ${f^{(i-1)}}''\neq {f^{(i)}}''$. We shall use the unit step
function $\mu$ for discontinuity of ${f^{(i)}}''$ at $t=t_i$.\vskip5pt

Consider the FRW metric of the form
$$
g=-dt^{2}+f^{2}(t)\left(\frac{dr^{2}}{1-kr^{2}}+r^{2}(d\theta^{2}+\sin^{2}\theta
d\phi^{2})\right) $$
where $k$ is a parameter denoting the
spatially flat ($k=0$), 3-sphere ($k=1$) and hyperboloid ($k=-1$) universes. \vskip10pt

{\bf Proposition 3.1} \  Let $M=M_0\times_{f} H$ be the FRW spacetime with Riemannian curvature $R$ and flow vector field $U=\partial_t$. If $f\in C^1(S)$, vector fields $X$, $Y$, $Z\in \frak{L}(H)$ satisfy
\begin{eqnarray}
&(i)&\ \ R_{XY}Z={\frac{{f'}^{2}+k}{f^{2}}}(\langle X,\ Z\rangle Y-\langle Y,\ Z\rangle X)\nonumber\\
&(ii)&\ \ R_{XU}U=\frac{{f}''}{f} X\nonumber\\
&(iii)&\ \ R_{XY}U=0\nonumber\\
&(iv)&\ \ R_{XU}Y=\frac{{f}''}{f}\langle X,\ Y\rangle U\nonumber
\end{eqnarray}
where $f^{\prime\prime}$ is given by
\begin{eqnarray}
f^{\prime\prime}&=&\left(f^{(n)\prime\prime}-f^{(n-1)\prime\prime}
+\frac{1}{n}\sum_{k=0}^{n-1}f^{(k)\prime\prime}\right)\mu(t-t_{n})\nonumber\\
&&+\sum_{l=1}^{n-1}\left(-f^{(l-1)\prime\prime}
+\frac{1}{n}\sum_{k=0}^{n-1}f^{(k)\prime\prime}\right)\mu(t-t_{l})\nonumber\\
&&+\frac{1}{n}\sum_{k=0}^{n-1}f^{(k)\prime\prime}\mu(t_{n}-t)
+\sum_{l=1}^{n-1}\left(-f^{(l)\prime\prime}
+\frac{1}{n}\sum_{k=0}^{n-1}f^{(k)\prime\prime}\right)\mu(t_{l}-t),
\label{fpp}
\end{eqnarray}
with $\mu(t-t_i)$ being the unit step function which becomes unity
for $t>t_{i}$ and vanishes otherwise. \vskip5pt

{\it{Proof.}}\par We derive $f^{\prime\prime}$ in terms of the
collection of functions $\{f^{(i)}\}$ with $f^{(i)}$ piecewisely
defined on the intervals $t_{i}\leq t\leq t_{i+1}$
$(i=0,1,2,...,n)$ with $t_{n+1}=t_{\infty}$.  For a single
discontinuity $n=1$ case, $f$ is trivially given by
$$
f^{\prime\prime}=f^{(1)\prime\prime}\mu(t-t_{1})+f^{(0)\prime\prime}\mu(t_{1}-t)
$$
which fulfills (\ref{fpp}). For double discontinuities $n=2$ case,
$f$ is similarly given by
\begin{eqnarray}
f^{\prime\prime}&=&\left(f^{(2)\prime\prime}-\frac{1}{2}f^{(1)\prime\prime}
+\frac{1}{2}f^{(0)\prime\prime}\right)\mu(t-t_{2})+\left(
\frac{1}{2}f^{(1)\prime\prime}-\frac{1}{2}f^{(0)\prime\prime}\right)\mu(t-t_{1})\nonumber\\
&&+\left(\frac{1}{2}f^{(1)\prime\prime}+\frac{1}{2}f^{(0)\prime\prime}\right)
\mu(t_{2}-t) +\left(-\frac{1}{2}f^{(1)\prime\prime}
+\frac{1}{2}f^{(0)\prime\prime}\right)\mu(t_{1}-t), \label{f2pp}
\end{eqnarray}
which also fulfills (\ref{fpp}). By using iteration method, one can
obtain (\ref{fpp}) for an arbitrary $n$ case. \hfill $\framebox{}$
\vskip10pt

For the case of $f\in C^0(S)$ we use the derivative of the unit step function $\mu(t_i)$. 
For all $t\not=t_i$ this is well-defined, $\mu'(t)=0$. However, at $t=t_i$ there exists a jump discontinuity so that we cannot define classical derivative and thus we use the $\delta$-function, $\mu'(t-t_i)=\delta(t-t_i)$ to obtain the follow results.\vskip10pt

{\bf Proposition 3.2}\  Let $M=M_0\times_{f} H$ be the FRW spacetime
with Riemannian curvature $R$ and flow vector field
$U=\partial_t$. If $f\in C^0(S)$, vector fields $X$, $Y$, $Z\in
\frak{L}(H)$ then satisfy
\begin{eqnarray}
&(i)&\ \ R_{XY}Z=\frac{f^{\prime 2}+k}{f^{2}}(\langle X,\ Z\rangle
Y-\langle Y,\ Z\rangle X)\nonumber\\
&(ii)&\ \ R_{XU}U=\frac{f^{\prime\prime}}{f} X\nonumber\\
&(iii)&\ \ R_{XY}U=0\nonumber\\
&(iv)&\ \ R_{XU}Y=\frac{f^{\prime\prime}}{f} \langle X,\ Y\rangle
U\nonumber
\end{eqnarray}
where $f^{\prime}$ and $f^{\prime\prime}$ are given by
\begin{eqnarray}
f^{\prime}&=&\left(f^{(n)\prime}-f^{(n-1)\prime}
+\frac{1}{n}\sum_{k=0}^{n-1}f^{(k)\prime}\right)\mu(t-t_{n})\nonumber\\
&&+\sum_{l=1}^{n-1}\left(-f^{(l-1)\prime}
+\frac{1}{n}\sum_{k=0}^{n-1}f^{(k)\prime}\right)\mu(t-t_{l})\nonumber\\
&&+\frac{1}{n}\sum_{k=0}^{n-1}f^{(k)\prime}\mu(t_{n}-t)
+\sum_{l=1}^{n-1}\left(-f^{(l)\prime}
+\frac{1}{n}\sum_{k=0}^{n-1}f^{(k)\prime}\right)\mu(t_{l}-t)\label{fp}\\
f^{\prime\prime}&=&\left(f^{(n)\prime\prime}-f^{(n-1)\prime\prime}
+\frac{1}{n}\sum_{k=0}^{n-1}f^{(k)\prime\prime}\right)\mu(t-t_{n})\nonumber\\
&&+\sum_{l=1}^{n-1}\left(-f^{(l-1)\prime\prime}
+\frac{1}{n}\sum_{k=0}^{n-1}f^{(k)\prime\prime}\right)\mu(t-t_{l})\nonumber\\
&&+\frac{1}{n}\sum_{k=0}^{n-1}f^{(k)\prime\prime}\mu(t_{n}-t)
+\sum_{l=1}^{n-1}\left(-f^{(l)\prime\prime}
+\frac{1}{n}\sum_{k=0}^{n-1}f^{(k)\prime\prime}\right)\mu(t_{l}-t)\nonumber\\
&&+\left(f^{(n)\prime}-f^{(n-1)\prime}\right)\delta(t-t_{n})
+\sum_{l=1}^{n-1}\left(f^{(l)\prime}-f^{(l-1)\prime}\right)\delta
(t-t_{l}), \label{fpp2}
\end{eqnarray}
with $\mu(t-t_i)$ and $\delta(t-t_{i})$ being the unit step
function and the delta function, respectively.\vskip5pt

{\it{Proof.}}\par Similar to (\ref{fpp}) in Proposition 3.1, one can
readily obtain $f^{\prime}$. Differentiating $f^{\prime}$ with
respect to $t$ and using the definition of the delta function
$\mu^{\prime}(t-t_{i})=\delta (t-t_{i})$ at $t=t_{i}$, one can
also obtain $f^{\prime\prime}$. \hfill $\framebox{}$ \vskip10pt

{\bf Proposition 3.3}\  Let $M=M_0\times_{f} H$ be the FRW spacetime
with Riemannian curvature $R$ and flow vector field
$U=\partial_t$. If $f\in C^0(S)$ and $X$, $Y\in
\frak{L}(H)$, then Ricci curvature is given by
\begin{eqnarray}
&(i)&\ {\text{Ric}}(U,U)=-{\frac{3{f}''}{f}} \nonumber\\
&(ii)&\ {\text{Ric}}(U,X)=0\nonumber\\
&(iii)&\ {\text{Ric}}(X,Y)=\left({\frac{2(f^{\prime
2}+k)}{f^2}}+{\frac{{f}''}{f}}\right)\langle X,\ Y\rangle,~~~~~
{\text{if}}\ \ X, Y\perp U\nonumber
\end{eqnarray}
where $f^{\prime}$ and $f^{\prime\prime}$ are given by (\ref{fp})
and (\ref{fpp2}). \vskip10pt

{\bf Proposition 3.4}\  Let $M=M_0\times_{f} H$ be the FRW spacetime
with Riemannian curvature $R$ and flow vector field
$U=\partial_t$. If $f\in C^0(S)$, the Einstein scalar
curvature is given by
$$
R=6\left(\frac{f^{\prime
2}}{f^{2}}+\frac{f^{\prime\prime}}{f}+\frac{k}{f^{2}}\right),
$$
where $f^{\prime}$ and $f^{\prime\prime}$ are given by (\ref{fp}) and (\ref{fpp2}).\vskip10pt

{\bf Proposition 3.5}\  For every plane containing a vector field of
$U=\partial_{t}$,  if $f\in C^0(S)$ and $X$, $Y\in
\frak{L}(H)$, we have a sectional curvature $K$ on the spacetime
$(M,\ g)$ for an arbitrary plane containing a vector field of $U=\partial_{t}$ and
$W={\alpha}U+{\beta}Y$
$$K(W,X)={\frac{-\alpha^2 {f}''+\beta^2\bigl({f}' +k\bigr)}{(-\alpha^2+\beta^2)f^2}}$$
where $f^{\prime}$ and $f^{\prime\prime}$ are given by
(\ref{fp}) and (\ref{fpp2}).\vskip10pt
{\it{Proof.}}\par The result is follows from $K(W,X)=\frac
{g(R_{WX}W, X)} {g(W, W)g(X,X)-[g(W, X)]^2}$ of the nondegenerate
2-plane with basis $(W, X)$.\hfill $\framebox{}$ \vskip10pt

\noindent{\bf IV.  Cosmology of spatially flat FRW metric with
double discontinuities}\vskip10pt
In the spatially flat FRW cosmology with $k=0$, the early universe
was radiation dominated, the adolescent universe was matter
dominated, and the present universe is now entering into
lambda-dominate phase in the absence of vacuum energy. If the
universe underwent inflation, there was a very early period when
the stress-energy was dominated by vacuum energy.  The Friedmann
equation may be integrated to give the age of the universe in
terms of present cosmological parameters.  We have the scale
factor $f$ as a function of time $t$ which scales as $f(t)\propto
t^{1/2}$ for a radiation-dominated (RD) universe, and scales as
$f(t)\propto t^{2/3}$ for a matter-dominated (MD) universe, and
scales as $f(t)\propto e^{Kt}$ for a lambda-dominated (LD)
universe.  Note that the transition from the radiation-dominated phase 
to the matter-dominated is not an abrupt one; neither is the later
transition from the matter-dominated phase to the exponentially
growing lambda-dominated phase. \vskip10pt
With the above astrophysical phenomenology in mind, consider the
spatially flat FRW spacetime $(M,\ g)$ with metric
$g=-dt^2+f^2(t)d\sigma^2$ in the form of warped products. Let
$M=M_0\times_{f} H$ be a warped product with
$g_{M_0}=-dt^2$.\vskip10pt

{\bf Definition  4.1}\ \ A $C^{0}$-Lorentzian metric on $M$ is a
nondegenerate (0,2) tensor of Lorentzian signature  such that\par
\hskip0.5cm$(i)$\ \ $g \in C^{0}$ on $S$\par 
\hskip0.45cm$(ii)$\ \ $g \in
C^{\infty}$ on $M\cap S^{c}$\par 
\hskip0.4cm$(iii)$\ \  for all $p \in
S$, and $U(p)$ partitioned by $S$, $g{\mid}_{U_{p}^{+}}$ and
$g{\mid}_{U_{p}^{-}}$ have smooth extensions to $U$. We call $S$ a
$C^{0}$-singular hypersurface of $(M, g)$.\vskip10pt

Consider $M_0 $ as a $C^{0}$-singular hypersurface of $(M, g)$. In
the spatially flat FRW spacetime, $f>0$ is smooth functions on
$M_0=(t_{0},\ t_{\infty})$ except at $t\not=t_{i}$ $(i=1,2)$, that
is $f\in C^{\infty}(S)$ (where $S=\{t_{i}\}\times_{f}H$) for
$t\not=t_{i}$ and $f\in C^{0}(S)$ at $t=t_{i}\in M_0$ to yield
\begin{equation}
f=\left(\begin{array}{l}
f^{(0)}=c_{0}t^{1/2},~~~~\mbox{for~$t<t_{1}$}\\
f^{(1)}=c_{1}t^{2/3},~~~~\mbox{for~$t_{1}\leq t\leq t_{2}$}\\
f^{(2)}=c_{2}e^{Kt},~~~~\mbox{for~$t>t_{2}$}
\end{array}
\right) \label{f012}
\end{equation}
with the boundary conditions
\begin{equation}
c_{0}t_{1}^{1/2}=c_{1}t_{1}^{2/3},~~~c_{1}t_{2}^{2/3}=c_{2}e^{Kt_{2}}.
\label{bc}
\end{equation}
Experimental values for $t_{1}$ and $t_{2}$ are given by
$t_{1}=4.7\times 10^{4}$ yr and $t_{2}=9.8$ Gyr~\cite{ry}.
Moreover $c_{1}$ and $c_{2}$ are given in terms of $c_{0}$,
$t_{1}$ and $t_{2}$ as follows
$$c_{1}=c_{0}t_{1}^{-1/6},~~~c_{2}=c_{0}t_{1}^{-1/6}t_{2}^{2/3}e^{-Kt_{2}}.$$
Note that in the spatially flat FRW model, $f\in C^0(S)$ since if
we assume $f\in C^1(S)$ one could have the boundary conditions
$\frac{1}{2}c_{0}t_{1}^{-1/2}=\frac{2}{3}c_{1}t_{1}^{-1/3}$ and
$\frac{2}{3}c_{1}t_{2}^{-1/3} =Kc_{2}e^{Kt_{2}}$, which cannot
satisfy the above boundary conditions (\ref{bc}) simultaneously.
\vskip10pt

{\bf Proposition 4.2}\  Let $M=M_0\times H$ be the spatially flat
FRW spacetime with Riemannian curvature $R$, flow vector field
$U=\partial_t$ and warping function $f\in C^0(S)$. For vector
fields $X$, $Y$, $Z\in \frak{L}(H)$ we have
\begin{eqnarray}
&(i)&\ \ R_{XY}Z={\frac{f^{\prime 2}}
{f^{2}}}(\langle X,\ Z\rangle Y-\langle Y,\ Z\rangle X)\nonumber\\
&(ii)&\ \ R_{XU}U={\frac{{f}''}{f}} X\nonumber\\
&(iii)&\ \ R_{XY}U=0\nonumber\\
&(iv)&\ \ R_{XU}Y={\frac{{f}''}{f}}\langle X,\ Y\rangle U\nonumber
\end{eqnarray}
where $f$ is given by (\ref{f012}) and $f^{\prime}$ and
$f^{\prime\prime}$ are given by
\begin{eqnarray}
f^{\prime}&=&\left(\frac{1}{4}c_{0}t^{-1/2}-\frac{1}{3}c_{1}t^{-1/3}
+Kc_{2}e^{Kt}\right)\mu(t-t_{2})\nonumber\\
&&+\left(-\frac{1}{4}c_{0}t^{-1/2}+\frac{1}{3}c_{1}t^{-1/3}
\right)\mu(t-t_{1})\nonumber\\
&&+\left(\frac{1}{4}c_{0}t^{-1/2}+\frac{1}{3}c_{1}t^{-1/3}\right)\mu(t_{2}-t)\nonumber\\
&&+\left(\frac{1}{4}c_{0}t^{-1/2}-\frac{1}{3}c_{1}t^{-1/3}\right)\mu(t_{1}-t)\label{fpfrw}\\
f^{\prime\prime}&=&\left(-\frac{1}{8}c_{0}t^{-3/2}+\frac{1}{9}c_{1}t^{-4/3}
+K^{2}c_{2}e^{Kt}\right)\mu(t-t_{2})\nonumber\\
&&+\left(\frac{1}{8}c_{0}t^{-3/2}-\frac{1}{9}c_{1}t^{-4/3}
\right)\mu(t-t_{1})\nonumber\\
&&+\left(-\frac{1}{8}c_{0}t^{-3/2}-\frac{1}{9}c_{1}t^{-4/3}\right)\mu(t_{2}-t)\nonumber\\
&&+\left(-\frac{1}{8}c_{0}t^{-3/2}+\frac{1}{9}c_{1}t^{-4/3}\right)\mu(t_{1}-t)\nonumber\\
&&+\left(-\frac{2}{3}c_{1}t^{-1/3}+Kc_{2}e^{Kt}\right)\delta (t-t_{2})\nonumber\\
&&+\left(-\frac{1}{2}c_{0}t^{-1/2}+\frac{2}{3}c_{1}t^{-1/3}\right)\delta
(t-t_{1}), \label{fppfrw}
\end{eqnarray}
with $\mu(t-t_i)$ and $\delta(t-t_{i})$ being the unit step
function and the delta function, respectively.\vskip5pt

{\it{Proof.}}\par Substituting $f$ in (\ref{f012}) into (\ref{fp})
and (\ref{fpp2}) in Proposition 3.2, one can readily obtain
(\ref{fpfrw}) and (\ref{fppfrw}). \hfill $\framebox{}$ \vskip10pt

{\bf Proposition 4.3}\  Let $M=M_0\times H$ be the spatially flat
FRW spacetime with Riemannian curvature $R$, flow vector field
$U=\partial_t$ and warping function $f\in C^0(S)$.  For vector fields 
$X$, $Y$, $Z\in \frak{L}(H)$, the Ricci curvature is given by
\begin{eqnarray}
&(i)&\ {\text{Ric}}(U,U)=-{\frac{3{f}''}{f}} \nonumber
\\ &(ii)&\ {\text{Ric}}(U,X)=0\nonumber
\\ &(iii)&\ {\text{Ric}}(X,Y)=\left({\frac{2f^{\prime
2}}{f^2}}+{\frac{{f}''}{f}}\right)\langle X,\
Y\rangle,~~~~~{\text{if}}\ \ X, Y\perp U\nonumber
\end{eqnarray}
where $f$, $f^{\prime}$ and $f^{\prime\prime}$ are given by
(\ref{f012}), (\ref{fpfrw}) and (\ref{fppfrw}),
respectively.\vskip10pt

{\bf Proposition 4.4}\ Let $M=M_0\times H$ be the spatially flat
FRW spacetime with Riemannian curvature $R$, flow vector field
$U=\partial_t$ and warping function $f\in C^0(S)$.  The Einstein scalar
curvature is then given by
$$
R=6\left(\frac{f^{\prime
2}}{f^{2}}+\frac{f^{\prime\prime}}{f}\right),
$$ where $f$, $f^{\prime}$ and $f^{\prime\prime}$ are given by
(\ref{f012}), (\ref{fpfrw}) and (\ref{fppfrw}),
respectively.\vskip10pt

{\bf Proposition 4.5}\  For every plane containing a vector field of
$U=\partial_{t}$ and $f\in C^0(S)$, if $X$, $Y\in
\frak{L}(H)$ we have a sectional curvature $K$ on the FRW spacetime $(M,\ g)$ 
for an arbitrary plane containing a vector field of $U=\partial_{t}$ and $W={\alpha}U+{\beta}Y$
$$K(W,X)={\frac{-\alpha^2 {f}''+\beta^2 {f}'} {(-\alpha^2+\beta^2)f^2}}$$
where $f$, $f^{\prime}$ and $f^{\prime\prime}$ are given by
(\ref{f012}), (\ref{fpfrw}) and (\ref{fppfrw}),
respectively.\vskip10pt
{\bf Proposition 4.6}\  Let $M=M_0\times H$ be the spatially flat
FRW spacetime with Riemannian curvature $R$, flow vector field
$U=\partial_t$ and warping function $f\in C^0(S)$.  The evolution equations
are then given by
\begin{eqnarray}
&&(i)\ \frac{3f^{\prime 2}}{f^{2}}=8\pi\rho+\Lambda\nonumber\\
&&(ii)\ \frac{3f^{\prime\prime}}{f}=-4\pi(\rho+3P)+\Lambda,
\nonumber
\end{eqnarray}
where $f$, $f^{\prime}$ and $f^{\prime\prime}$ are given by
(\ref{f012}), (\ref{fpfrw}) and (\ref{fppfrw}), respectively. Here
$\rho$, $P$ and $\Lambda$ are the mass density and pressure of
matter and the cosmological constant. \vskip10pt

{\it{Proof.}}\par Consider the Einstein equation
\begin{equation}
G_{\mu\nu}+\Lambda
g_{\mu\nu}=R_{\mu\nu}-{\frac{1}{2}}Rg_{\mu\nu}+\Lambda
g_{\mu\nu}=8\pi T_{\mu\nu} \label{eeqn} \end{equation} where
$G_{\mu\nu}$ is the Einstein tensor, and $T_{\mu\nu}$ is the
stress-energy tensor for all the field present-matter, radiation
and so on.  To be consistent with the symmetries of the metric,
the total stress-energy tensor $T_{\mu\nu}$ must be diagonal, and
by isotropy the spatial components must be equal. The simplest
realization of such a stress-energy tensor is that of a perfect
fluid characterized by a time-dependent energy density $\rho(t)$
and pressure $p(t)$,

\begin{equation}
T^{\mu}_{\nu}={\text{diag}}(\rho,-p,-p,-p). \label{tmunu}
\end{equation}
Substituting (\ref{tmunu}) into (\ref{eeqn}), together with the
Ricci and Einstein curvatures given in Proposition 4.3 and Proposition 4.4, 
one can readily obtain the above evolution equations. \hfill
$\framebox{}$ \vskip10pt

{\bf Remarks 4.7}\  The $\mu=0$ component of the conservation
of stress-energy tensor, $T^{\mu\nu}_{;\nu}=0$, gives the first law of
thermodynamics of the familiar form $d(\rho f^3)=-pd(f^3)$ or equivalently, $d[f^3(\rho+p)]=f^3dp$.  
The change in energy in a co-moving volume element, $d(\rho f^3)$,
is equal to minus the pressure times the change in volume element,
$-pd(f^3)$. For the simple equation of state $p=\omega \rho$,
where $\omega$ is independent of time, the energy density evolves
as $\rho\propto f^{-3(1+\omega)}$. Examples of interest include: radiation 
$(p={\frac{1}{3}}\rho$, $\rho\propto f^{-4}$), matter $(p=0$, $\rho\propto f^{-3})$ 
and vacuum energy $(p=-\rho$, $\rho\propto$ const.) phases.\vskip10pt

\noindent{\bf V. Conclusions}\vskip10pt

We have considered the FRW cosmological model in the warped
product scheme to investigate the non-smooth curvature associated
with the multiple discontinuities involved in the evolution of the
universe. In particular we have analyzed the non-smooth features
of the spatially flat FRW universe by introducing double
discontinuities occurred at the radiation-matter and matter-lambda
phase transitions in the astrophysical phenomenology.\vskip10pt

\begin{center}
{\bf Acknowledgments}
\end{center}

JC and STH would like to acknowledge financial support in part
from the Korea Science and Engineering Foundation Grant
(R01-2001-000-00003-0) and (R01-2000-00015).


\begin{thebibliography}{99}

\bibitem{guth} A.H. Guth, The inflationary universe: a possible solution to the
horizon and flatness problems, Phys. Rev. D23, 347-356 (1981).
\bibitem{seiberg} J. Khoury, B.A. Ovrut, N. Seiberg, P.J. Steinhardt and N. Turok,
From big crunch to big bang, Phys. Rev. D65, 086007 (2002).
\bibitem{rn} H. Reissner, \"Uber die eigengravitation des elektrischen felds nach der
Einsteinshen theorie, Ann. Phys. 50, 106-120 (1916); G. Nordstr\"om, On the energy of the gravitational 
field in Einstein's theory, Proc. Kon. Ned. Akda. Wet. 20, 1238-1245
(1918). 1238-1245.
\bibitem{ksy} J. Demers, R. Lafrance, and R.C. Meyers, Black hole entropy without brick walls, Phys. Rev. D52, 2245-2253 (1995); A. Ghosh and P. Mitra, Entropy for extremal Reissner-Nordstrom black holes, Phys. Lett. B357, 295-299 (1995); S.P. Kim, S.K. Kim, K.S. Soh and J.H. Yee, Remarks on renormalization of black hole entropy, Int. J. Mod. Phys. A12, 5223-5234 (1997); G. Cognola and P. Lecca, Electromagnetic fields in Schwarzschild and Reissner-Nordstrom geometry, Phys. Rev. D57, 1108-1111 (1998).
\bibitem{choi00} J. Choi, Multiply warped products with nonsmooth metrics, J. Math. Phys. 41, 8163-8169 (2000); S.T. Hong, J. Choi and Y.J. Park, (2+1) dimensional black holes in warped products, Gen. Rel. Grav. 35 (2003), in press {\tt gr-qc/0209058}; S.T. Hong, J. Choi and Y.J. Park, Warped products and Reissner-Nordstrom metric, {\tt math.DG/0204273}.
\bibitem{fs} J.L. Flores and M. S\'{a}nchez, Geodesic connectedness of multiwarped spacetimes, J. Diff. Eqn. 
186, 1 (2002). 
\bibitem{li} A. Lichnerowicz, {\it Theor\'{e}or\'{i}es relativistes de la gravitation et de l'\'{e}lectromag\'{e}ticame} (Masson et Cie, Paris, 1955).
\bibitem{sm1} J. Smoller and B. Temple, Shock waves near the Schwarzchild radius and stability limits for stars, Phys.Rev. D55, 7518-7528 (1997).
\bibitem{sm2} J. Smoller and B. Temple, Cosmology with shock wave, Comm. Math. Phys. 210, 275-308 (2000). 
\bibitem{gk12} R.F. Hoskins, {\it Generalized Functions} (Ellis Horwood limited, Oxford, 1979). 
\bibitem{ta} A.N. Taylor and P.I.R. Watts, Evolution of the cosmological density distribution function, {\tt astro-ph/0001118}.
\bibitem{bo} R.L. Bishop and B. O'Neill, Manifolds of negative curvature, Trans. Amer. Math. Soc. 145, 1 (1969).
\bibitem{bep} J.K. Beem, P.E. Ehrlich and T.G. Powell, Warped product manifolds in relativity, in selected studies: 
{\it A Volume Dedicated to the Memory of Allbert Einstein} (North-Holland, Amsterdam, 41, 1982) Eds., T.M. 
Rassias and G. M. Rassias. 
\bibitem{on} B. O'Neill, {\it Semi-Riemannian Geometry with Applications to Relativity} (Academic Press 
Pure and Applied Mathematics, 1983).      
\bibitem{ry} B. Ryden, {\it Introduction to cosmology} (Addison Wesley, New York, 2003).
\end{thebibliography}
\end{document}